\documentclass[conference]{IEEEtran}
\IEEEoverridecommandlockouts
\normalsize

\ifCLASSINFOpdf

\else

\fi
\usepackage{soul}
\usepackage{array}
\usepackage{color}
\usepackage{amsfonts}
\usepackage{amssymb}
\usepackage{amsmath}
\usepackage{graphicx}
\usepackage{mathtools}
\usepackage{cite}
\usepackage{balance}
\usepackage{caption}
\usepackage{subcaption}
\usepackage{textcomp}
\usepackage{gensymb}
\usepackage{float}
\usepackage{comment}
\usepackage{subcaption}
\usepackage{listings}
\usepackage{tabularx}
\usepackage{multirow}
\usepackage{amsthm}
\usepackage{algorithmic}
\usepackage{algorithm}

\newcommand\given[1][]{\:#1\vert\:}
\usepackage{bm}
\usepackage{fancyhdr}
\usepackage[utf8]{inputenc}
\begin{document}
\title{Deep Learning Model-Based Channel Estimation for THz Band Massive MIMO with RF Impairments}
%\author{Author 1, Author 2, Author 3,  Author 4, and Author 5}
\author{\IEEEauthorblockN{Pulok Tarafder$^1$, Imtiaz Ahmed$^1$, Danda B. Rawat$^1$, Ramesh Annavajjala$^2$, and Kumar Vijay Mishra$^3$}\\\IEEEauthorblockA{$^{1}$Howard University, Washington, DC, USA\\$^2$University of Massachusetts, Boston, MA, USA\\$^3$United States DEVCOM Army Research Laboratory, Adelphi, MD, USA}}

\maketitle
\thispagestyle{fancy}
\begin{abstract}

THz band enabled large scale massive MIMO (M-MIMO) is considered as a key enabler for the 6G technology, given its enormous bandwidth and for its low latency connectivity. In the large-scale M-MIMO configuration, enlarged array aperture and small wavelengths of THz results in an amalgamation of both far field and near field paths, which makes tasks such as channel estimation for THz M-MIMO highly challenging. Moreover, at the THz transceiver, radio frequency (RF) impairments such as phase noise (PN) of the analog devices also leads to degradation in channel estimation performance. Classical estimators as well as traditional deep learning (DL) based algorithms struggle to maintain their robustness when performing for large scale antenna arrays i.e., M-MIMO, and when RF impairments are considered for practical usage. To effectively address this issue, it is crucial to utilize a neural network (NN) that has the ability to study the behaviors of the channel and RF impairment correlations, such as a recurrent neural network (RNN). The RF impairments act as sequential noise data which is subsequently incorporated with the channel data, leading to choose a specific type of RNN known as bidirectional long short-term memory (BiLSTM) which is followed by gated recurrent units (GRU) to process the sequential data. Simulation results demonstrate that our proposed model outperforms other benchmark approaches at various signal-to-noise ratio (SNR) levels.
\end{abstract}


\section{Introduction}
With the recent advancements in sixth-generation (6G) communication, it has been identified that spectrum bottleneck is a key limitation of enabling increased data rates \cite{dovelos2021channel}. To facilitate the overgrowing data demands, the new terahertz (THz) band from 0.1 to 10 THz has been identified as the key enabler to ensure extreme data rates \cite{cai2024toward, elbir2021terahertz}. However, one of the core challenges in THz band is its high attenuation, which is caused by high spreading loss and molecular absorption \cite{sha2021channel}. In response, massive multiple-input multiple-output (M-MIMO) technology has emerged as a promising solution, leveraging the simultaneous utilization of hundreds of antennas at the base station to overcome the challenges such as path loss and blockage \cite{li20145g}. Moreover, compared to millimeter-wave (mmWave) frequencies, THz channels are extremely sparse and typically modeled as line-of-sight (LoS)-dominant \cite{elbir2023federated}. THz band occupies a spectral domain that bridges the mmWave frequencies and the infrared band, and it presents a unique challenge for signal generation across both electronic and photonic devices, resulting in radio-frequency (RF) impairments \cite{kallfass201564,kallfass2015towards,grzyb2018high,rodriguez201916}.

Electromagnetic (EM) radiation field can typically be categorized into far-field and near-field regions, and it is expected that both of these components will be exploited in future 6G mobile networks. As a result, in the THz enabled M-MIMO systems literature, there are typically two categories of low-overhead channel estimation schemes are present, i.e., far-field channel estimation \cite{wei2020deep,gao2019wideband,lee2016channel} and near-field channel estimation \cite{han2020channel}. For far-field channel estimation, the channel sparsity is considered in the angle domain, where signals can only be pointed towards 
a specific direction \cite{zhang20236g}. Whereas, near-field channel estimation considers aperture arrays will experience spherical wavefronts \cite{yin2017scatterer}, and the channel sparsity is in polar domain. 

With a few exceptions where deep learning (DL) methodologies were used, most studies in the literature that addressed far-field, near-field, or hybrid-field ultra M-MIMO configurations \cite{hu2022hybrid} focused on traditional channel estimating techniques. In \cite{lee2016channel}, an orthogonal matching pursuit (OMP) is used to estimate the far-field angular-domain channel. In \cite{han2020channel}, Han \textit{et al.} exploited dedicated sparsity patterns and OMP for near-field estimation. Cui \textit{et al.} in \cite{cui2022channel} proposed a polar-domain simultaneous OMP (PSOMP) channel estimator for near-field communication. Furthermore, a hybrid-field OMP based channel estimator was proposed in \cite{wei2021channel}. A fixed point theory based DL assisted channel estimator is presented in \cite{yu2022hybrid} for a hybrid-field channel model. Another DL based channel estimation algorithm is presented in \cite{zhang2023near} while focusing only on near-field channel model. In \cite{lei2023channel}, a novel near-field channel estimation algorithm is presented using residual dense networks. 

To the best of our knowledge, no prior work on hybrid-field channel estimation techniques have addressed RF impairment (phase noise (PN)) incorporated intelligent DL aided channel estimation algorithm for THz ultra M-MIMO (UMMIMO) communication systems. In this paper, owing to the inherent challenges constituted by the PN and the small wavelengths of high frequency THz-band, we propose a DL assisted intelligent channel estimation architecture for hybrid-field THz band UMMIMO architecture. Although our proposed scheme is constructed for THz band, the scheme can also be utilized in very high frequency mmWave band. Our proposed method captures the dynamic behaviors of PN in the hybrid-field channel model and accurately estimates the channel utilizing the powerful capabilities of bidirectional long short-term memory (BiLSTM) and gated recurrent units (GRU) combined. BiLSTM exploits the advantage of the sequential prediction tasks combined with GRU to accelerate the training process. 

The remainder of the paper is organized as follows. Section \ref{Section2} describes the THz-band system model. Section \ref{Section3} discusses the proposed DL framework. Afterwards, section \ref{Section4} presents the performance analysis of the proposed framework, and some concluding remarks are summarized in section \ref{Section5}. 
 
\section{System Model}\label{Section2}
Within the framework of UMMIMO systems functioning in the THz-band, we consider a base station (BS) constructed with an ultra-large-scale array of $N$ antenna components. The objective of using a large antenna array at BS is to facilitate downlink communication with a single-antenna user equipment (UE). Let $\Hat{\Hat{\bm{H}}} \in \mathbb{C}^{1 \times N}$ denote the hybrid channel characterizing the propagation environment between the BS and the UE. To model the downlink channel estimation process, we consider the following signal representation:
\begin{equation}
\boldsymbol{y} = \Hat{\Hat{\bm{H}}} \boldsymbol{\Phi} + \boldsymbol{w}, \label{eq1Sys}
\end{equation}
where $\boldsymbol{y} \in \mathbb{C}^{1 \times M}$ represents the received signal observations at the UE over $M$ pilot vectors, $\boldsymbol{\Phi} \in \mathbb{C}^{N \times M}$ denotes the known pilot matrix transmitted by the BS, and $\boldsymbol{w}$ $\in \mathbb{C}^{1 \times M}$ denotes independent and identically distributed (i.i.d.) complex additive white Gaussian noise (AWGN) with zero mean and variance $\sigma^2$. Note that \eqref{eq1Sys} does not incorporate PN in the considered model. We leverage PN models for both transmitter and receiver and update the system model at the end of this section.

Given the knowledge of $\boldsymbol{y}$ and $\boldsymbol{\Phi}$, the downlink channel estimation problem entails the recovery of the channel vector $\Hat{\Hat{\bm{H}}}$. In the context of UMMIMO systems, the number of antennas ${N}$ at the BS is substantially large. In these systems, the spatial dynamics of signal transmission necessitate a comprehensive model that can simultaneously account for the different characteristics of far-field and near-field propagation. To this end, for our analysis, we utilize an advance channel model known as hybrid-field channel model \cite{cui2022channel}.

\begin{figure}[t!]
\centerline{\includegraphics[width=\linewidth]{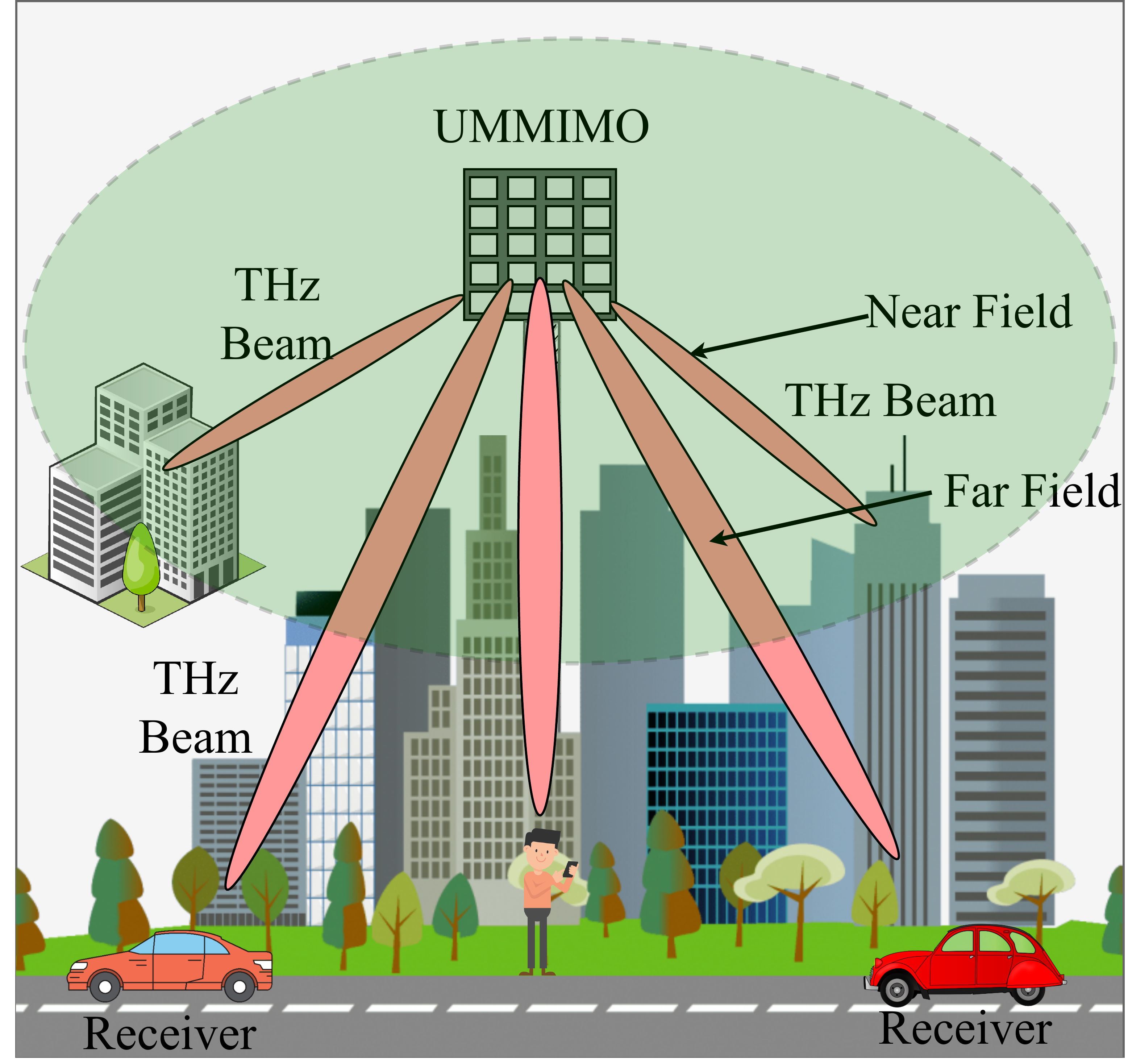}}
\caption{THz Massive MIMO.}
\label{fig1}
\end{figure}

% \begin{figure}[t]
% \centerline{\includegraphics[width=\linewidth]{Figures/near_far.png}}
% \caption{Far-Field and Near-Field Illustration.}
% \label{fig2}
% \end{figure}

% \begin{figure*}[t!]
% %\centerline{\includegraphics[scale=0.65]{system_model.pdf}}
% \centerline{\includegraphics[scale=0.48]{Figures/System Architecture.pdf}}
% \caption{Proposed deep learning model with BiLSTM-GRU.}
% \label{fig3}
% \end{figure*}
\begin{figure*}[t!]
    \centering
    \includegraphics[width=\textwidth]{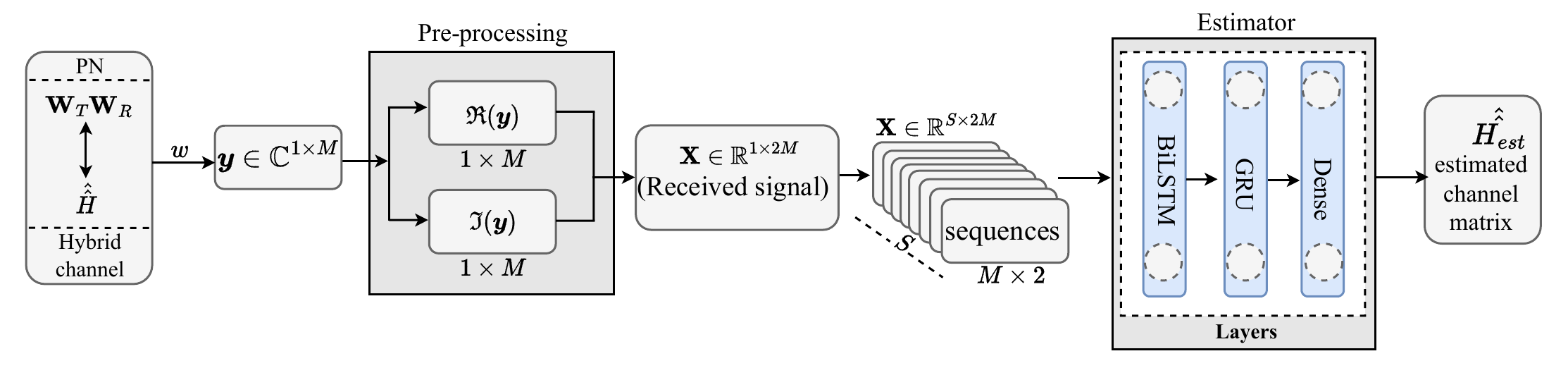}
    \caption{Proposed deep learning model with BiLSTM-GRU.}
    \label{fig3}
\end{figure*}

\subsection{Far-Field Component}

The far-field component of the channel $\Hat{\Hat{\bm{H}}}$ can be modeled as a sum of $L_f$ plane waves impinging on the antenna array from different angles. The far-field channel component, denoted by $\boldsymbol{H}_f \in \mathbb{C}^{1 \times N}$, can be expressed as:
\begin{equation}
   \boldsymbol{H}_f = \sum_{l=1}^{L_f} \alpha_f^{(l)} \mathbf{a}_f(\theta_f^{(l)}),
\end{equation}
where $L_f$ is the number of far-field paths, $\alpha_f^{(l)} \in \mathbb{C}$ denotes the complex gain of the $l^{th}$ far-field path, $\theta_f^{(l)}$ is the angle of arrival of the $l^{th}$ far-field path, and $\mathbf{a}_f(\theta_f^{(l)}) \in \mathbb{C}^{1 \times N}$ is the far-field array steering vector, given by:

% \begin{equation}
%    \mathbf{a}_f(\theta_f) = \frac{1}{\|\mathbf{a}_f'(\theta_f^{(l)})\|}  
%    \begin{bmatrix}
%        1 \\
%        e^{j\pi\sin(\theta_f^{(l)})} \\
%        \vdots \\
%        e^{j\pi(N-1)\sin(\theta_f^{(l)})}
%    \end{bmatrix}
% \end{equation}

\begin{equation}
   \mathbf{a}_f(\theta_f) = \frac{1}{\|\mathbf{a}_f'(\theta_f^{(l)})\|}
   \left[1, e^{j\pi\sin(\theta_f^{(l)})}, \ldots, e^{j\pi(N-1)\sin(\theta_f^{(l)})}\right]^T.
\end{equation}

\subsection{Near-Field Component}

The near-field component of the $\Hat{\Hat{\bm{H}}}$ can be modeled as a sum of $L_{nf}$ spherical wavefronts originating from different distances and angles. The near-field channel component, denoted by $\boldsymbol{H}_{nf} \in \mathbb{C}^{1 \times N}$, can be expressed as:
\begin{equation}
   \boldsymbol{H}_{nf} = \sum_{l=1}^{L_{nf}} \alpha_{nf}^{(l)} \mathbf{a}^l_{nf}(\Delta_{nf}^{(l)}, \theta_{nf}^{(l)}),
\end{equation}
where $L_{nf}$ is the number of near-field paths, $\alpha_{nf}^{(l)}$ presents the complex gain of the $l^{th}$ near-field path, $\Delta_{nf}^{(l)}$ denotes the distance of the $l^{th}$ near-field path, $\theta_{nf}^{(l)}$ depicts the angle of arrival of the $l^{th}$ near-field path, and $\mathbf{a}_{nf}(\Delta_{nf}^{(l)}, \theta_{nf}^{(l)})$ is the near-field array steering vector, given by:
% \begin{equation}
% \mathbf{a}_{nf}(\Delta_{nf}^{(l)}, \theta_{nf}^{(l)}) = \frac{1}{\Lambda}
% \scalebox{1.2}{    \renewcommand{\arraystretch}{1.5} % Increase the row spacing
%     \begin{bmatrix}
%    \displaystyle\frac{\Delta_{nf}^l}{\xi_{nf}^{(1)}} e^{j\frac{2\pi f}{c} \left(\xi_{nf}^{(1)} - \Delta_{nf}^{(l)}\right)} \\
%    \displaystyle\frac{\Delta_{nf}^l}{\xi_{nf}^{(2)}} e^{j\frac{2\pi f}{c} \left(\xi_{nf}^{(2)} - \Delta_{nf}^{(l)}\right)} \\
%    \vdots \\
%    \displaystyle\frac{\Delta_{nf}^l}{\xi_{nf}^{(N)}} e^{j\frac{2\pi f}{c} \left(\xi_{nf}^{(N)} - \Delta_{nf}^{(l)}\right)}
%     \end{bmatrix}
%     }
% \end{equation}
% \\
\begin{multline}
    \mathbf{a}_{n f}\left(\Delta_{n f}^{(l)}, \theta_{n f}^{(l)}\right) = \frac{1}{\Lambda} \left[\frac{\Delta_{n f}^{(l)}}{\xi_{n f}^{(1)}} e^{j \frac{2 \pi \digamma}{c} \left(\xi_{n f}^{(1)} - \Delta_{n f}^{(l)}\right)}, \right.\\
    \left. \frac{\Delta_{n f}^{(l)}}{\xi_{n f}^{(2)}} e^{j \frac{2 \pi \digamma}{c} \left(\xi_{n f}^{(2)} - \Delta_{n f}^{(l)}\right)},\ldots, \frac{\Delta_{n f}^{(l)}}{\xi_{n f}^{(N)}} e^{j \frac{2 \pi \digamma}{c} \left(\xi_{n f}^{(N)} - \Delta_{n f}^{(l)}\right)}\right]^T.
\end{multline}
% \begin{equation}
% \mathbf{a}_{n f}\left(\Delta_{n f}^{(l)}, \theta_{n f}^{(l)}\right) = \frac{1}{\Lambda} \left[ \begin{array}{@{\quad}c@{\quad}}
% \scalebox{1.1}{$\displaystyle\frac{\Delta_{n f}^{(l)}}{\xi_{n f}^{(1)}} e^{j \frac{2 \pi \digamma}{c} \left(\xi_{n f}^{(1)} - \Delta_{n f}^{(l)}\right)}$} \\[2em]
% \scalebox{1.}{$\displaystyle\frac{\Delta_{n f}^{(l)}}{\xi_{n f}^{(2)}} e^{j \frac{2 \pi \digamma}{c} \left(\xi_{n f}^{(2)} - \Delta_{n f}^{(l)}\right)}$} \\[1em]
% \vdots \\[1em]
% \scalebox{1.1}{$\displaystyle\frac{\Delta_{n f}^{(l)}}{\xi_{n f}^{(N)}} e^{j \frac{2 \pi \digamma}{c} \left(\xi_{n f}^{(N)} - \Delta_{n f}^{(l)}\right)}$}
% \end{array} \right]
% \end{equation}\\
Here $\Lambda = \|\mathbf{a}_{nf}'(\Delta_{nf}^{(l)}, \theta_{nf}^{(l)})\|$ and $c$ is the speed of light. Moreover, $\digamma$ is the carrier frequency, $\Delta_{nf}^{(l)}$ denotes the distance from the $l^{th}$ scatter to the center of the antenna array, and $\xi_{nf}^{(k)}\given[\Big]_{k=1}^N$ depicts the distance from the $k^{th}$ antenna element to the corresponding scatter. Furthermore, $\xi_{nf}^{(k)}\given[\big]_{k=1}^N$ can be  expressed by:

\begin{equation}
   \xi_{nf}^{(k)}\given[\big]_{k=1}^N = \sqrt{{(\Delta_{nf}^{(l)})}^2 + \left(\Psi_{k} d \right)^2 - 2 \Delta_{nf}^{(l)} \Psi_{k} d \sin(\theta_{nf}^{(l)})},
\end{equation}
where $\Psi_k = \frac{2k - N - 1}{2}$ and $d$ denotes the antenna spacing.

\subsection{Hybrid Channel Model}
% The initial hybrid channel model, denoted by $\hat{H} \in \mathbb{C}^{1 \times N}$, is a superposition of the far-field and near-field components can be expressed as:
In the hybrid-field model, the two propagation methodologies are distinguished by Rayleigh distance $\Omega=\frac{2D^2}{\lambda}$, where $D$ is the antenna array aperture of radiation, and $\lambda$ represents the wavelength. Note that $\Omega$ is the key parameter, which determines whether the channel acts as near field or far field. Near-field channel exists at a region $\Theta < \Omega$ whereas far-field exists at $\Theta \geq \Omega$. In particular,
\[
\text{Channel Region}(\Theta) = 
\begin{cases} 
\text{Near-field}, & \text{if } \Theta < \Omega \\
\text{Far-field}, & \text{if } \Theta \geq \Omega
\end{cases}.
\]
The comprehensive hybrid channel model, encapsulated within $\Hat{\bm{H}} \in \mathbb{C}^{1 \times N}$, emerges from the combination of the far-field and near-field components. This model is formulated through the equation:
% \begin{equation}
%    \hat{H} = \boldsymbol{H}_f \given[\big]_{\{\Theta > \Omega\}} + \boldsymbol{H}_{nf} \given[\big]_{\{\Theta < \Omega\}}
% \end{equation}
\begin{equation}
   \Hat{\bm{H}} = \sum_{L_f=1}^{\gamma L} \boldsymbol{H}_f \given[\big]_{\{\Theta \geq \Omega\}} + \sum_{L_{nf}=1}^{(1-\gamma)L} \boldsymbol{H}_{nf} \given[\big]_{\{\Theta < \Omega\}},
\end{equation}
where $\gamma \in [0,1]$ and $L$ indicate the number of possible pathways. The crucial factor in determining the number of far-field and near-field components in the propagation environment is $\gamma$. The considered hybrid-field channel model reduces to the conventional far-field channel model for $\gamma = 1$. On the other hand, it turns into a near-field channel model when $\gamma = 0$. In order to guarantee a consistent overall power, the hybrid channel is normalized as follows:
\begin{equation}
   \Hat{\Hat{\bm{H}}} =  \Hat{\bm{H}} \sqrt{\frac{N}{L_f + L_{nf}}}.
\end{equation}
% \subsection{Phase Noise}
% Once we have the channel, we incorporate RF impairments known as PN in this context. First, we denote $\theta_R[n]:$ as receiver Rx PN and $\theta_T[n]:$ as transmitter Tx PN. Afterwards, we finalize the incorporation by introducing both of the PNs with the channel model as follows

% \begin{equation}\label{eq6}
% \boldsymbol{y}[p]=e^{-j \theta_R\lfloor p \rfloor}\left(\Hat{\Hat{H}}[p] *\left(\boldsymbol{\Phi}[p] e^{j \theta_T\lfloor p]}\right)\right)+\boldsymbol{w}[p].
% \end{equation}

% \noindent\textbf{Problem Formulation:} 

\subsection{Phase Noise Incorporation}
In the domain of high-frequency M-MIMO systems, the manifestation of accurately modeling RF impairments becomes complicated owing to PN being a predominant factor that significantly affects system performance. PN, originating from the inherent instability in the oscillator frequencies at both the transmitter (Tx) and receiver (Rx), manifests as random phase variations over time. This section delineates the mathematical framework adopted to integrate the PN into the hybrid channel model, thereby capturing the essence of real-world RF imperfections at THz-bands.

Let us denote $\theta_T$ and $\theta_R$ as the PN processes at the Tx and Rx, respectively. For each sample $n$, the Wiener process random-walk \cite{mehrpouyan2012joint} PN for the Tx and Rx can be represented as:
\begin{equation}
\begin{aligned}
    \theta_{T}[n] &= \theta_{T}[n-1] + \Delta\theta_{T}[n]~\text{and}\\ \theta_{R}[n] &= \theta_{R}[n-1] + \Delta\theta_{R}[n].
\end{aligned}
\end{equation}
% Here, $\Delta\theta_{T}[n] \sim \mathcal{N}(0,\sigma_{T}^2)$ and $\Delta\theta_{R}[n] \sim \mathcal{N}(0,\sigma_{R}^2)$ both are normally distributed ($\mathcal{N}$) with mean $0$ and variance $\sigma^2$.
Here, $\Delta\theta_{T}[n] \sim \mathcal{N}(0,\sigma_{T}^2)$ and $\Delta\theta_{R}[n] \sim \mathcal{N}(0,\sigma_{R}^2)$ represent changes in PN for the Tx and Rx, respectively. Each is normally distributed ($\mathcal{N}$) with a mean of $0$ and variances $\sigma_{T}^2$ for the Tx and $\sigma_{R}^2$ for the Rx. The mathematical representation of the received signal, encompassing the influence of PN alongside impact on the channel is expressed as follows:
% \begin{equation}\label{eq10}
% \boldsymbol{y}=e^{-j \theta_R \rfloor}\left(\Hat{\Hat{H}}[p] *\left(\boldsymbol{\Phi}[p] e^{j \theta_T\lfloor p]}\right)\right)+\boldsymbol{w}[p]
% \end{equation}
\begin{equation}\label{eq9}
\boldsymbol{y} = e^{-j \theta_R}\left(\Hat{\Hat{\bm{H}}}\boldsymbol{\Phi} e^{j \theta_T}\right) + \boldsymbol{w}.
\end{equation}
Here, $e^{j\theta_T}$ and $e^{-j\theta_R}$ represent the phase modulation and demodulation processes attributable to the Tx and Rx PNs, respectively. Note that, Eq.~(\ref{eq9}) can be further be simplified as:
\begin{equation}\label{eq10}
\boldsymbol{y} =\mathbf{W}_T \mathbf{W}_R \Hat{\Hat{\bm{H}}} \boldsymbol{\Phi} + \boldsymbol{w},
\end{equation}
where, the symbol $\boldsymbol{y}$ represents the received signal, and $\mathbf{W}_T \in \mathbb{R}^{1 \times N}$ and $\mathbf{W}_R \in \mathbb{R}^{N \times 1}$ represent the Tx and Rx PN components, respectively.
%%%%%%

\section{Proposed LSTM-based PN Channel Estimation}\label{Section3}
% This section describes a Long Short-Term Memory (LSTM) based approach for channel estimation in the presence of PN. Traditional channel estimation methods often struggle to handle the dynamic nature of PN. Our proposed LSTM-based approach leverages the sequential learning capabilities of LSTMs to effectively capture the temporal dependencies introduced by PN. The goal is to feed $\boldsymbol{y}$ into the model and predict $\mathbf{W}_T \mathbf{W}_R \Hat{\Hat{H}}$ combined, considering PN $\mathbf{W}_T \mathbf{W}_R$ and channel $\Hat{\Hat{H}}$ as a joint channel properties under the assumption that $\boldsymbol{\Phi}$ is known.

This section describes a BiLSTM and GRU based channel estimation scheme for the considered system. Neural networks (NNs) excel at classification and recognition tasks where the output is a discrete label, and information loss within the model may not significantly impact performance. However, for data generation problems such as channel estimation, where the output is a continuous signal or waveform, information loss can lead to degraded performance \cite{dong2020channel}. In addition, the dynamic nature of PN \cite{rasekh2021phase} incorporated with the channel makes the channel estimation task more complicated. Conventional feed forward deep NN (DNN) based channel estimation methods do not have the capability to exploit the long term channel correlations. Techniques that prioritize information preservation are more suitable than conventional NN models optimized for classification or recognition. Our proposed BiLSTM-based approach leverages the sequential learning capabilities of LSTMs to effectively capture the temporal dependencies introduced by PN. The goal is to accurately estimate the combined effect of PN ($\mathbf{W}_T \mathbf{W}_R$) and the channel characteristics ($\Hat{\Hat{\bm{H}}}$). This proposed scheme considers the influence of PN and channel properties as intertwined elements of a singular, comprehensive channel model. It is worth mentioning that the pilot signal matrix $\boldsymbol{\Phi}$ is pre-designed and accurately known to both Tx and Rx.

\subsection{Data Preprocessing}
Before feeding the data into the LSTM model, a crucial preprocessing is performed for efficient training. Note that we consider a downlink transmission model with received signal processing on the UE side. At the UE, the complex received signals $\boldsymbol{y} \in \mathbb{C}^{1 \times M}$ are separated into its real and imaginary components at first, then we concatenated the real and imaginary parts to construct the input dataset for the proposed scheme. Given the received signal vector set denoted by $\boldsymbol{y} \in \mathbb{C}^{1 \times M}$, the separation and concatenation of the real and imaginary components can be mathematically represented as follows.

Let $\Re(\boldsymbol{y} \in \mathbb{C}^{1 \times M})$ denote the real part and $\Im(\boldsymbol{y} \in \mathbb{C}^{1 \times M})$ denote the imaginary part of the complex channel phenomena. The concatenated vector for each sample then becomes:
\begin{equation}
\mathbf{X} = [\Re(\boldsymbol{y} \in \mathbb{C}^{1 \times M}),  \Im(\boldsymbol{y} \in \mathbb{C}^{1 \times M})] \in \mathbb{R}^{1 \times 2M},
\end{equation}
where $\mathbf{X}$ represents the new data structure after separation and concatenation of the real and imaginary components. For $S$ number of training samples, this transformation is applied to each sample, resulting in the dimensionality transformation as follows:
\begin{equation}
\boldsymbol{y} \in \mathbb{C}^{1 \times M} \in \mathbb{C}^{S \times M} \Rightarrow \mathbf{X} \in \mathbb{R}^{S \times 2M}.
\end{equation}

\subsection{Proposed Model Architecture}
% In the construction of our proposed model as portrayed in Fig.~\ref{fig3}, we have architected a DL network leveraging the robustness of BiLSTM and GRU, followed by a dense layer. LSTM is a promising denoiser in the context of M-MIMO channel estimation at higher frequencies \cite{pulok2023lstm}. In the LSTM module, hidden layers has the advantage of accessing information from the past inputs. Nevertheless, in the challenging scenario of hybrid-field THz channel estimation, characterized by extreme noise levels, PN, and signal attenuation, BiLSTM emerges as a superior choice. Its architecture, which facilitates the processing of input data in both forward and reverse temporal directions, enables a comprehensive analysis of the signal's past and future states. This dual-directional processing capability makes BiLSTM exceptionally adept at capturing the dynamic fluctuations of PN inherent in highly noisy and attenuated THz communication environments. Afterwards, GRU compensates and retrieves the data lost during the denoising by BiLSTM.

In the construction of our proposed model as portrayed in Fig.~\ref{fig3}, we design a DL network leveraging the robustness of BiLSTM and GRU, followed by a dense layer. LSTM is a promising denoiser in the context of channel estimation for UMMIMO at higher frequencies \cite{pulok2023lstm}. The hidden layers of the LSTM cell can capture the important information from the past and avoid the redundant information, thus providing a greater ability to capture the information compared to the simple RNN cell. The structure of the BiLSTM network is the combination of two LSTM networks with two different directions \cite{schuster1997bidirectional}. Therefore, the BiLSTM approach facilitates the processing of input data in both forward and reverse temporal directions, and enables a comprehensive analysis of the past and future states of the signal. This dual-directional processing capability makes BiLSTM exceptionally adept at capturing the dynamic fluctuations of PN inherent in highly noisy and attenuated THz communication environments. Afterwards, GRU compensates and retrieves the data lost during the denoising operation by BiLSTM.

In the DL module, the model expects an input sequence of dimension $2N$. 
This input shape is reshaped to $2N \times 1$ for processing, implying that each sequence consists of $2N$ timesteps, with a single feature per timestep for the BiLSTM and GRU layers. The single feature at each timestep represents the concatenated real-valued signal vector for a particular Tx-Rx antenna pair. Both the BiLSTM and GRU layers are designed to process this input structure, each equipped with $N$ LSTM and GRU cells respectively to ensure the consistent flow of information throughout the network. Finally, the dense layer with a single neuron and a linear activation function concludes the proposed DL network and maps the high-level learned features from the previous two layers and predicts a single continuous value from the sequence. It is worth noting that we trained our proposed model offline using comprehensive datasets, and subsequently evaluated its performance on live network conditions online.

\begin{table}[t]
\caption{Adopted simulation parameters} 
\centering
\renewcommand{\arraystretch}{1.8}

\begin{tabular}{|l|l|}
\hline

\textbf{Hyperparameter} & \textbf{Value} \\
\hline

Linear signal-to-noise ratio (SNR)  & $1/{\sigma^2}$ \\
\hline
SNR (dB) range & $[0:5:20]$ \\
\hline
BS antenna array $N$ & $64, 128$ \\
\hline
Total number of paths $L$  & $4$ \\
\hline
Hybrid-field path component $\gamma$ & $0.5$ \\
\hline
Sub-THz carrier frequency $\digamma$  & $100$ GHz \\
\hline
$\Delta_{nf}^{(l)}$min & $10~m$\\
\hline
$\Delta_{nf}^{(l)}$max & $80~m$\\
\hline
Speed of light $c$ & $3\times10^8~ms^{-1}$\\
\hline
Initial $\sigma_{T}^2$ & $0.1$\\
\hline
Initial $\sigma_{R}^2$ & $0.2$\\
\hline
\end{tabular}
\label{table1}
\end{table}

\section{Simulation Results}\label{Section4}
This section includes a comprehensive numerical study of our proposed DL framework and an assessment of its performance in relation to conventional channel estimation schemes. We create the training datasets for our simulation using Eq.~(\ref{eq10}). Additionally, Table~\ref{table1} provides a summary of all the simulation parameters. In this work, we employ the normalized mean square error (NMSE) as the primary performance assessment metric to evaluate the accuracy of our channel estimation model at various signal-to-noise ratio (SNR) levels, where we define SNR as $1/{\sigma^2}$. The NMSE is used to quantify the deviation between the ground truth channel matrix $\Hat{\Hat{\bm{H}}}$ and the estimated channel matrix $\Hat{\Hat{\bm{H}}}_{est}$ obtained from our model. A lower NMSE value indicates a smaller deviation between the estimated channel matrix and the ground truth, signifying a more accurate channel estimation. The NMSE estimate in decibel (dB) scale is calculated by
\begin{equation} \label{NMSE}
    \text{NMSE} = 10 \log \left\{ {\mathbb{E}} \left [ \frac{||\Hat{\Hat{\bm{H}}} - \Hat{\Hat{\bm{H}}}_{est}||^2}{||\Hat{\Hat{\bm{H}}}||^2} \right ] \right \},
\end{equation}
where $\mathbb{E}$ denotes the statistical expectation.

\begin{figure}[t!]
\centerline{\includegraphics[width=\linewidth]{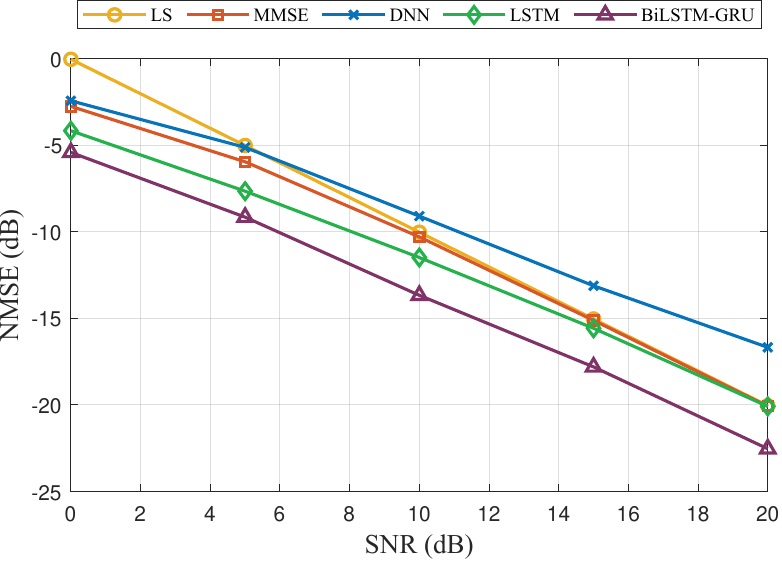}}
\caption{NMSE vs. SNR (dB) for $N=64$.}
\label{fig3}
\end{figure}

\begin{figure}[h!]
\centerline{\includegraphics[width=\linewidth]{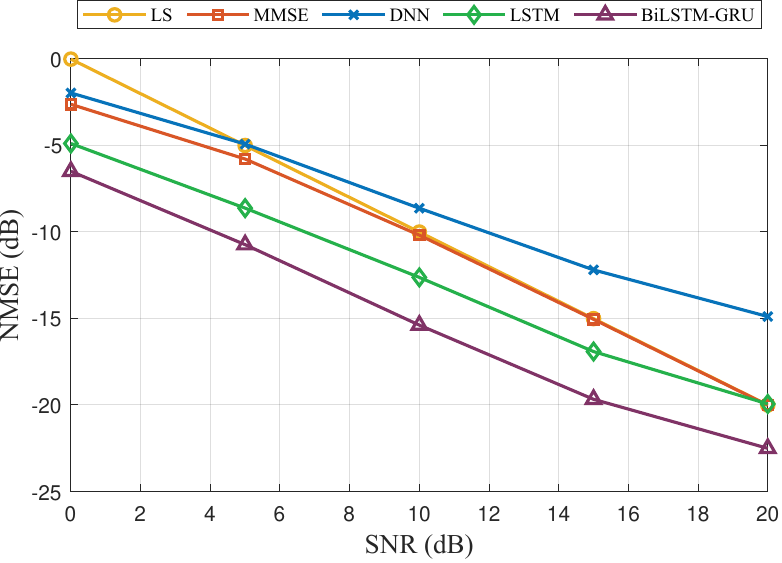}}
\caption{NMSE vs. SNR (dB) for $N=128$.}
\label{fig4}
\end{figure}

\begin{figure}[t!]
\centerline{\includegraphics[width=\linewidth]{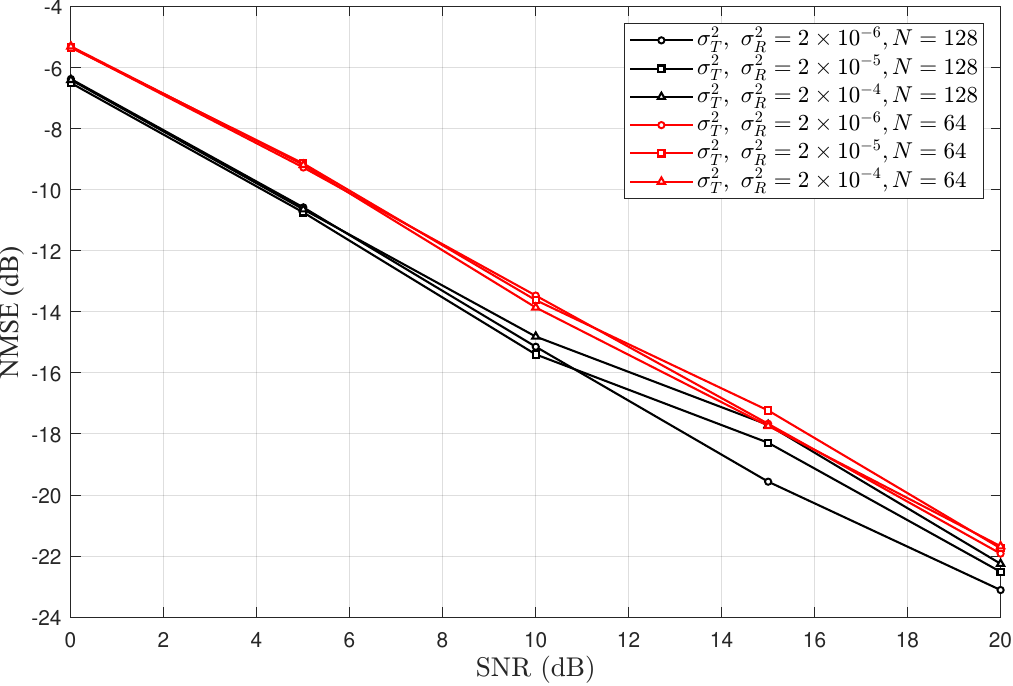}}
\caption{Performance Analysis.}
\label{fig5}
\end{figure}

% \subsection{Simulation Results}
At first, using Eq.~(\ref{eq10}), we generate datasets for $N = 64$ and $N = 128$ separately, each having $6000$ data samples, for various SNR levels i.e., $0, 5, 10, 15$, and $20$ dB. We then divide the generated dataset into 80\% for training and 20\% for testing. We utilize TensorFlow \cite{abadi2016tensorflow} backend Keras to execute the DL simulation. We train each dataset separately using Adam optimizer \cite{kingma2017adam}. We tune the model with $0.001$ learning rate and batch size $16$, and set mean square error (MSE) as the loss function until convergence, exploiting the callbacks feature of TensorFlow. We then compare the performance of our proposed DL model based scheme with the conventional least square (LS) and minimum mean square error (MMSE) estimation schemes, DNN, and LSTM. Fig.~\ref{fig3} and \ref{fig4} demonstrate the NMSE performance comparison against LS, MMSE, DNN, and LSTM at various SNR levels, and Fig.~\ref{fig5} presents the performance of our DL model at multiple levels of PN variance. We plotted NMSE values at five different levels stated above and observed that our proposed scheme performs very consistently and outperforms other algorithms by large margin.

In Fig.~\ref{fig3}, for $N = 64$, it is portrayed that our proposed model achieves an estimation NMSE of $-5.42$ dB at $0$ dB SNR, whereas are LS obtains $-0.04$ dB, MMSE obtains $-2.76$ dB, DNN obtains $-2.438$ dB, and standalone LSTM obtains $-4.167$ dB. On the other hand, at $20$ dB SNR level, they obtain $-20.05$ dB, $-20.08$ dB, $-16.679$ dB, and $-20.089$ dB respectively, and our proposed model achieves $-22.538$ dB. In this scenario, at the very low SNR, our model outperforms LS, MMSE, DNN, and LSTM by $99\%$,~$49\%$,~$55\%$, and $23\%$ margins respectively, whereas at very high SNR, it outperforms by $11\%,~10\%,~25\%$, and $10.8\%$ margins respectively in terms of NMSE. 

We can observe the similar trend in estimation performance for $N = 128$ in Fig.~\ref{fig4}. At low SNR, our model achieves $-6.511$ dB whereas LS, MMSE, DNN, and LSTM obtains $0.03$ dB,  $-2.64$ dB, $-1.984$ dB, and $-4.9037$ dB respectively. It is evident that the proposed scheme outperforms LS, MMSE, DNN, and LSTM by $99.5\%,~ 59.45\%,~69.52\%,$ and $24.68\%$ respectively in terms of NMSE. Similarly, at very high SNR, e.g., $20$ dB, we outperform LS, MMSE, DNN, LSTM by $11.11\%,~11.07\%,~33.81\%,$ and $11.39\%$ respectively.

Furthermore, in Fig.~\ref{fig5} we also compare the performance of our model at various 
$\sigma_{T}^2$ and $\sigma_{R}^2$ for both $N = 128$ and $64$. As we can observe from the figure that it compares the performance under three different phase noise variances, i.e. $2 \times 10^{-6}$, $2 \times 10^{-5}$, and $2 \times 10^{-4}$. Higher PN adversely impacts the stability and accuracy of the signal processing at both the Tx and Rx. For all configurations, the NMSE decreases with increasing SNR. This trend is consistent and expected, as higher SNR implies a stronger signal relative to noise, leading to more accurate channel estimation. In addition, with a larger number of antennas, $(N = 128)$ the channel estimation task yields consistently better performance than with fewer antennas $(N = 64)$.

\section{Conclusions}\label{Section5}
In this paper, we have proposed a novel BiLSTM-GRU based channel estimation algorithm for high frequency UMMIMO networks that operate in the THz-band. The proposed model is designed and deployed with a combination of BiLSTM, GRU, and Dense layers. The first BiLSTM layer performs the denoising of the channel and GRU layers accelerates the training while smoothing the channel matrices. We have compared our proposed algorithm with baselines LS, MMSE, LSTM, and DNN and furthermore have evaluated the performance in terms of NMSE at various PN noise variances. The results concluded that our model is robust in both high and low SNR scenarios and outperforms the baseline consistently by a large margin.

%\bibliographystyle{IEEEtran}
%\bibliography{references_TeamB}
\balance
\bibliographystyle{IEEEtran}
\bibliography{references_TeamB}
\balance

\end{document}